\documentclass[conference,a4paper]{IEEEtran}
\IEEEoverridecommandlockouts
\usepackage{cite}
\usepackage{amsthm,amsmath,dsfont}
\usepackage{algorithmic}
\usepackage{textcomp}
\usepackage{xcolor}

\usepackage{comment}

\usepackage{float}
\usepackage{subfig}
\usepackage{wrapfig}
\usepackage{framed,xcolor}
\usepackage{newfloat}
\usepackage[xcolor,framemethod=TikZ]{mdframed}
\usepackage{amsthm}
\usepackage{thmtools}
\usepackage{stmaryrd}
\usepackage{mathtools}
\usepackage{lineno}

\usepackage{graphicx}
\usepackage{amssymb,xfrac}
\usepackage{thmtools}
\allowdisplaybreaks

\usepackage[algoruled,titlenotnumbered]{algorithm2e}
\usepackage{algorithmic}
\usepackage{wrapfig}
\usepackage{comment}
\usepackage{multirow}
\usepackage{amsmath,bm}
\usepackage{amsfonts,amssymb,dsfont} %%%%Lin
\usepackage{tikz}
\usepackage[utf8]{inputenc}
\usepackage{array}
\usepackage{makecell}

\usetikzlibrary{shapes.geometric,arrows,shapes,chains,calc,patterns,decorations.pathreplacing}

\newcommand{\EE}{\mathbb{E}}

\newcommand{\PP}{\mathbb{P}}

\newcommand{\A}{\mathrm{A}}
\newcommand{\B}{\mathrm{B}}

\newcommand{\tabincell}[2]{\begin{tabular}{@{}#1@{}}#2\end{tabular}}

\newcommand{\mathleft}{\@fleqntrue\@mathmargin0pt}

\usepackage[a4paper,bindingoffset=0.2in,%
left=0.514in,right=0.5in,top=1.53in,bottom=0.75in,%
footskip=.25in]{geometry}

\begin{document}

\renewcommand{\baselinestretch}{1.0}

\title{Comparative Analysis of Economic Instruments in Intersection Operation: A User-Based Perspective\\
%{\footnotesize \textsuperscript{*}%Note: Sub-titles are not captured in Xplore and should not be used
%}
%\thanks{This research was funded by the New York University Abu Dhabi Research Institute.
%}
}
\author{%Li~Li,~%\IEEEmembership{Fellow,~OSA,}
	Dianchao~Lin$^{\dagger}$,~%\IEEEmembership{Fellow,~OSA,}
	Saif~Eddin~Jabari,~%,~\IEEEmembership{Life~Fellow,~IEEE}% <-this % stops a space
	\thanks{This work was supported by the NYUAD Center for Interacting Urban Networks (CITIES), funded by Tamkeen under the NYUAD Research Institute Award CG001 and by the Swiss Re Institute under the Quantum Cities™ initiative.}
	%\thanks{Li Li is with the Department of Civil and Urban Engineering, New York University Tandon School of Engineering, NY, USA (email: ll3252@nyu.edu)}
	\thanks{Saif Eddin Jabari is with the Division of Engineering, New York University Abu Dhabi, UAE and the Department of Civil and Urban Engineering, New York University Tandon School of Engineering, NY, USA (email: sej7@nyu.edu)}
	\thanks{$^{\dagger}$ Corresponding author. Dianchao Lin is with the Department of Civil and Urban Engineering, New York University Tandon School of Engineering, NY, USA (email: dl3404@nyu.edu)}
}

\begin{comment}

\author{\IEEEauthorblockN{DianChao Lin}
\IEEEauthorblockA{\textit{Department of Civil and Urban Engineering} \\
\textit{New York University Tandon School of Engineering}\\
NY, USA \\
email: dl3404@nyu.edu}

\and
\IEEEauthorblockN{Saif Eddin Jabari} %2\textsuperscript{nd} 
\IEEEauthorblockA{\textit{Department of Civil and Urban Engineering} \\
	\textit{New York University Tandon School of Engineering}\\
%	\setlength{\baselineskip}{20pt}
	NY, USA \\ 
	\\
	%\setlength{\baselineskip}{12pt}
\textit{Division of Engineering} \\
\textit{New York University Abu Dhabi}\\
Abu Dhabi, UAE \\
email: sej7@nyu.edu}
}
\end{comment}
\maketitle

\begin{abstract}
Focusing on different economic instruments implemented in intersection operations under a connected environment, this paper analyzes their advantages and disadvantages from the travelers' perspective. Travelers' concerns revolve around whether a new instrument is easy to learn and operate, whether it can save time or money, and whether it can reduce the rich-poor gap. After a comparative analysis, we found that both credit and free-market schemes can benefit users. Second-price auctions can only benefit high VOT vehicles. From the perspective of technology deployment and adoption, a credit scheme is not easy to learn and operate for travelers.
%This document is a model and instructions for \LaTeX.
%This and the IEEEtran.cls file define the components of your paper [title, text, heads, etc.]. *CRITICAL: Do Not Use Symbols, Special Characters, Footnotes, 
%or Math in Paper Title or Abstract.
\end{abstract}

\begin{IEEEkeywords}
connected vehicle, economic instrument, auction, transaction, credit, intersection, operation, benefit, user-based
%component, formatting, style, styling, insert
\end{IEEEkeywords}

%\section{Structure of Paper}
%Market mechanism in microscopic traffic operation

%1 Introduction
%2 Comparative Analysis
%2.1 Convenience \& Complexity
%2.2 Flexibility
%2.3 Economic Benefit 
%3 Computational Personal Benefit
%3.1 Case 1: No abandonment and honest VOT reporting
%3.2  Case 2: Abandonment with dishonest VOT reporting
%3.3  Case 3: No abandonment wish dishonest VOT reporting
%4 Numerical Calculation of Benefit
%4.1 Pre-setting
%4.2 personal Benefit of Case 1
%4.3 personal benefit of Case 2
%4.4 personal benefit of Case 3
%5 Conclusion

\section{Introduction}

Economic instruments are known to be powerful traffic management tools, both theoretically and in practice (e.g. dynamic tolls). Under limited road resources, economic instruments with demand management can better serve heterogeneous vehicles' demand and improve overall social welfare. Although Pigou has numerically argued that pricing strategies are useful for improving social welfare in 1920 \cite{pigou2017economics}, their real-world implementation has not been possible until today. Inconvenience is one of the reasons that impede its actual deployment. Traditional tolls (e.g., toll booths) introduce delays to travelers. For example, paying for parking will result in no less than 10-60 seconds time lost to every vehicle. Thanks to RFID technology, tolling on the highway can be carried out with zero delays to travelers. RFID cards and gates can save people's time substantially. However, tolling gates and other facilities are usually expensive, and it has yet not seen wide-spread implementation, especially in many developing countries.

Today, we are witnessing major shifts in surface transportation. Advanced information technology and connected vehicle (CV) technology allows for direct vehicle-to-vehicle and vehicle-to-infrastructure communications, while auto-pilot features allow the vehicles to perform driving maneuvers autonomously (e.g., in response to the controls).  This has given rise to several recent papers to use tolling or auctions directly between vehicles for deciding on right-of-way (ROW) or priority \cite{schepperle2007agent,schepperle2008auction,vasirani2012market, carlino2013auction}. Auctions can be thought of as advanced tolling schemes, whereby vehicles auction their time in ways that capture heterogeneity in value of time (VOT).  Some have been used for intersection operation research and parking management research \cite{schepperle2007agent, schepperle2008auction, vasirani2012market, carlino2013auction, levin2015intersection,mashayekhi2015multiagent,raphael2017intersection,molinari2018automation, molinari2018traffic,lin2019transferable}.
% If the length of paper is too short, we can cite network tolling papers here 
In the intersection control literature, auctions require vehicles to give prices for a priority pass, and only vehicle(s) with a higher price(s) can pay and pass with priority. There are two typical formats: open auctions and sealed bid auctions
\cite{krishna2009auction}.  Open auctions include ascending price auctions and descending price auctions; sealed-bid auctions include first-price auctions, second-price auctions, etc. Sealed bid auctions are more suitable for traffic management without unduly burdening
drivers.

%In addition to tolling and auctions, there are two other main types of market mechanisms (mainly used in network traffic management research) that can also be used in microscopic traffic management. These are \emph{tradable credit schemes} \cite{yang2011managing,nie2012transaction} and \emph{tradable permit schemes} \cite{akamatsu2017tradable}. %Both are used in traffic network management research. 
%Tradable credit is similar to a new currency, which can only be used for travel.  By controlling the amount of credit in circulation, transportation managers can control traffic flow indirectly. The purpose of tradable permit schemes is making every section/intersection's passing permit be unique, hence traffic managers can easily control traffic flow directly.
Standard auction mechanisms are not \emph{revenue-neutral} economic instruments: travelers always need to pay an extra amount of money to travel. It can worsen family finances, especially for the poor, and unavoidably lead to public outcry and political opposition. The \emph{credit auction schemes}, which may be originated from \emph{tradable network credits theory} in traffic network management \cite{yang2011managing,nie2012transaction}, can also be revenue-neutral \cite{carlino2013auction}. The credit is similar to a new currency, which should be pre-distributed by the government freely and periodically. Citizens can use it for travel and trade it in a trading market. By controlling the amount of credit in circulation\footnote{In traffic network management, the government can also control required credit for passing a link}, the transport sector can also manage the generation of traffic demand indirectly. %The purpose of tradable permit schemes is making every section/intersection's passing permit be unique, hence traffic managers can easily control traffic flow directly. 

However, credit trading is itself time consuming and it renders travel a more complex process. A recent departure from auction-based mechanisms was the \emph{free-market mechanism} introduced in \cite{lin2019pay,lin2019transferable,lin2019pay1,henry2019free,lin2020pay}, in which vehicles \emph{pay each other} for right-of-way as opposed to paying a third party (the system operator).  Unlike auction-based approaches, these direct transaction-based approaches have the advantage that losers are directly compensated for ``giving'' priority. 

%These mechanisms are hard to popularize. Tradable credit and tradable permits can be revenue-neutral, but they require a separate trading market for credits/permits and a distribution program. This comes at extra costs to the users and results in a deviation from the desired network equilibrium state \cite{nie2012transaction}. Also, both tradable mechanisms require an initial distribution of credits/permits periodically. Such a distribution is expensive and hard to implement, especially when considering other practical aspects such as population mobility among cities, equity issues, and opportunities to game the system. The proposed model can operate without an external trading or auction system, and it does not require the distribution of initial credits/permits.

%To some degree, tradable credits/permits are also auction-based mechanisms. 
%A recent departure from auction-based mechanisms was proposed by us in \cite{lin2019pay,lin2019transferable}, in which vehicles \emph{pay each other} for right-of-way as opposed to paying a third party (the system operator).  Unlike auction-based approaches, these direct transaction-based approaches have the advantage that losers are compensated for ``giving'' priority. 

All above mechanisms of economic instruments, including auction, credit(credit auction), and direct-transaction(free-market),  are not easy to understand for citizens. Their concerns are usually different from and more complicated than the optimal objectives of theoretical models. Investigations are showing that travelers' major concerns include relative convenience, cost, speed, reliability, safety, comfort, environment pollution, body health, and so on \cite{jones2012motivations, eker2020exploratory}. We suppose that different economic instruments have a similar influence on safety, comfort, reliability, environment and health.  Among other things, they may care about: 
1) whether a new measure is easy to learn and operate;
2) whether it can save time or money; and even 
3) whether it is equity. Improvement of social welfare (even a Pareto-improvement\footnote{an improvement only benefit some people with the other people unchanged}) may also widen the rich-poor gap.
Focusing on the first and the second concerns, this paper comparatively analyzes the influence of different economics instruments for travelers.

The next section discusses what citizens need to do for their travel under different economic instruments. 
The following section analyzes citizens' benefits under different financial instruments. We transfer the time-saving options into monetary income and use the benefit to represent the sum income (transferred and real). 
This is followed by the last section, which briefly discusses the rich-poor gap and concludes the paper.

\section{Manual operation for travelers}
\label{S:mo}

%Implement a new economic instrument usually requires an extra manual operational process for travelers. 
Compared with auctions and direct-transactions, use of credit schemes require additional credit trading work (\ref{SS:tc}). Also, all systems need to trade for priority at the intersection (\ref{SS:tp}), and CV technology with the support of mobile payment is also necessary. Some steps of the process can be replaced by a machine, but other steps cannot. %It means that manual operation cannot be avoided.

\subsection {Trade of credit in virtual market}
\label{SS:tc}

The pre-distributed credit from the government will typically not be suitable in that the budget does not correctly align with their travel needs. Trade of credit cannot be avoided. %Otherwise, travelers cannot decide or change their travel plans by themselves. 

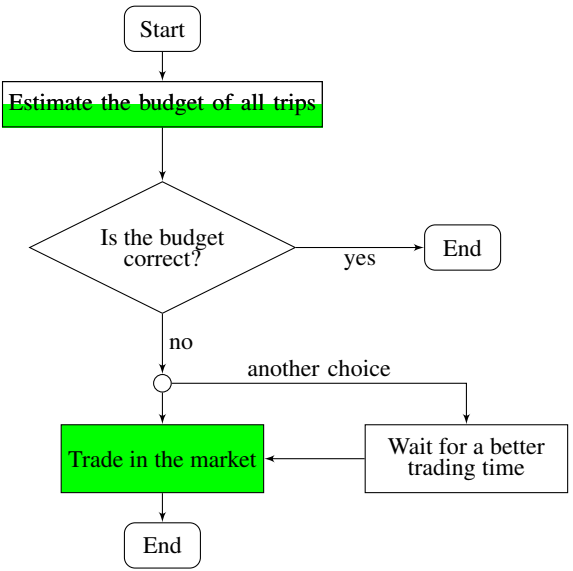
\begin{figure}[h!]
	\scriptsize
	\tikzstyle{startstop} = [rectangle, rounded corners, draw,thin,fill=white,align=center,text centered,minimum width=1cm,minimum height=0.6cm]%minimum width=3cm, minimum height=1cm,text centered] %, draw=black, fill=red!30
	\tikzstyle{format}=[rectangle,draw,thin,align=center,minimum width=1cm,minimum height=0.6cm]
	%	\tikzstyle{fill lower half} = {path picture={\fill[#1] (path picture bounding box.south west)
	%	rectangle (path picture bounding box.east);
	\begin{center}
		\begin{tikzpicture}[node distance=2cm,
		auto,>=latex',
		thin, 
		start chain=going below,
		every join/.style={norm},]
		\node (start) [startstop] {\small Start};
		%\fill[green]    (-3,-1) rectangle  ++ (3,-0.5);
		%	\draw         (-3,-1) rectangle  ++ (3,0);
		
		\node[below of=start,format,yshift=1cm](A){\small Estimate the budget of all trips};%\mycbox{green}
		\fill[green](A.south west) rectangle (A.east);
		\node[below of=start,format,yshift=1cm](A1){\small Estimate the budget of all trips};%\mycbox{green}
		\node [draw, diamond, aspect=2,below of=A,text width=2cm,align=center,yshift=0.1cm] (B){\small Is the budget correct?};
		\node (end1) [below of=A,xshift=3.95cm,yshift=0.1cm,startstop] {\small End};
		\node [below of=B,yshift=0.1cm,circle, draw,thin,fill=white,align=center,text centered,yshift=0.1cm,minimum width=0.2cm,minimum height=0.2cm](C){};
		\node[below of=C,format,yshift=1cm,text width=2.5cm,minimum height=0.9cm,fill=green](D){\small Trade in the market};
		\node[below of=C,format,xshift=4cm,yshift=1cm,text width=2.5cm,minimum height=0.9cm](E){\small Wait for a better trading time};
		\node (end2) [below of=D,yshift=0.85cm,startstop] {\small End};
		\draw[->] (start.south) -- (A.north);
		\draw[->] (A.south) -- (B.north);
		\draw[->] (B.east) -- node[anchor=north] {\small yes} (end1.west);
		\draw[->] (B.south) -- node[anchor=west] {\small no} (C.north);
		\draw[->] (C.south) -- (D.north);
		\draw[->] (C.east) --node[anchor=south] {\small another choice}(4,-4.7) -- (E.north);
		\draw[->] (D.south) -- (end2.north);
		\draw[->] (E.west) -- (D.east);
		\end{tikzpicture}
	\end{center}
	\caption{Flow chart of trading credit budget.}\label{F:fc1}
\end{figure}

Fig. \ref{F:fc1} illustrates the basic process of trading credit budget. Suppose the result of the step "Estimate the budget of all trips" is almost correct. The process in Fig. \ref{F:fc1} will be employed in two scenarios. 1) It happens at the beginning of credit refreshment (old credits are scraped, and new credits are distributed). 2) It starts if a traveler adds/cancels a trip in his/her travel schedule. It is hard to say which scenario is more likely, but it is clear that: if the period of credit redistribution is shorter, scenario one will happen more frequently; otherwise, scenario two will happen more frequently.% because of the uncertainty of the future.

The color green represents the work of the step that can be replaced/assisted by a machine. For example, we can imagine that the credits in different cities are similar to different stocks\footnote{The credits should be homogeneous within the same traffic network, such as the network of a city}. Travelers can know the prices and easily trade credits by mobile payment. 

\begin{figure}[h!]
	\scriptsize
	\tikzstyle{startstop} = [rectangle, rounded corners, draw,thin,align=center,text centered,minimum width=1cm,minimum height=0.6cm]%minimum width=3cm, minimum height=1cm,text centered] %, draw=black, fill=red!30 %fill=white,
	\tikzstyle{format}=[rectangle,draw,thin,align=center,minimum width=1cm,minimum height=0.6cm]%
	\begin{center}
		\begin{tikzpicture}[node distance=2cm,
		auto,>=latex',
		thin,
		start chain=going below,
		every join/.style={norm},]
		\node (start) [startstop] {\small Start};
		\node[below of=start,format,xshift=-2.3cm,yshift=0.5cm](A){\small VOT};
		\node[below of=start,format,yshift=0.5cm](B){\small Time schedule};
		\node[below of=start,format,xshift=2.5cm,yshift=0.5cm](C){\small OD pair};
		\node[below of=B,format,yshift=0.5cm,fill=green](D){\small Pick a route};
		\node[below of=D,format,yshift=1cm,fill=green](F){\small Estimate the required credits in this trip};
		\node (end) [below of=F,yshift=1cm,startstop] {\small End};
		\draw[->] (start.south) -- (B.north);
		\draw[->] (start.south) -- (0,-0.75) -- (-2.3,-0.75) -- (A.north);
		\draw[->] (start.south) -- (0,-0.75) -- (2.5,-0.75) -- (C.north);
		\draw[->] (B.south) -- (D.north);
		\draw[->] (A.south) -- (-2.3,-2.25) -- (0,-2.25) -- (D.north);
		\draw[->] (C.south) -- (2.5,-2.25) -- (0,-2.25) -- (D.north);
		\draw[->] (D.south) -- (F.north);
		\draw[->] (F.south) -- (end.north);
		\end{tikzpicture}
	\end{center}
	\caption{Flow chart of credits estimation in one trip.}\label{F:fc2}
\end{figure}
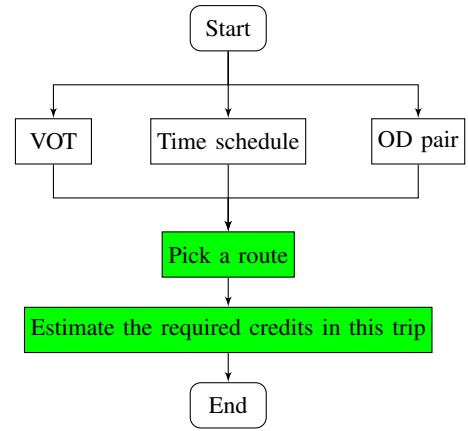

The step "estimate the budget of all trips" is a recombined step, and it can be described by the process in Fig. \ref{F:fc2}. If there are $n$ trips, this process in Fig. \ref{F:fc2} should be done $n$ times. At the beginning of Fig. \ref{F:fc2}, the input of travel information cannot be done by a machine, and travelers have to decide and provide such inputs themselves.%himself/herself.

%It is worth mentioning that, there are three reasons to lead to the fluctuation of credit's price. 1) The feature of credit. The newly updated credit is valuable and can be used during the whole period, but the credit will devalue quickly at the end of a period. 2) The uncertainty of travel. There are many emergencies that can significantly influence the travel in a city, such as the spread of diseases and the terrorist attack. Even a normal event, such as an unexpected change of weather, can influence the travel a lot. The change of the supply and demand relationship will also make price change. 3) The behavior and choice of people. It is easy to plan the commuting travel, but it is hard to predetermine the other travels, especially for the travel in the weekend. Also, in order to coping with uncertainty (required credit and best price), travelers may buy or sell credit more than they need to do. It can further influence the price of credit.

\subsection {Trade for priority at intersections}
\label{SS:tp}

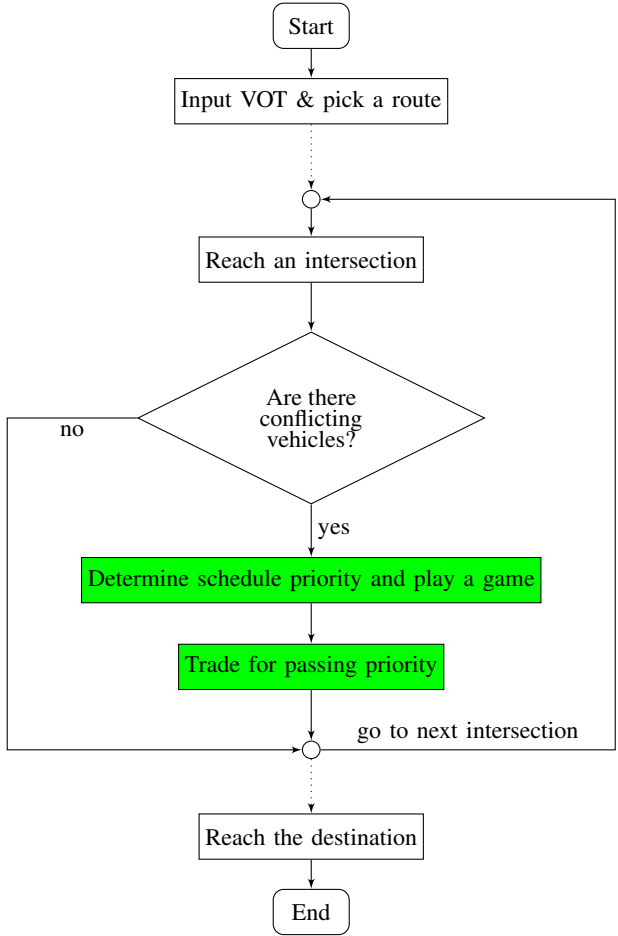
\begin{figure}[h!]
	\scriptsize
	\tikzstyle{startstop} = [rectangle, rounded corners, draw,thin,align=center,text centered,minimum width=1cm,minimum height=0.6cm]%minimum width=3cm, minimum height=1cm,text centered] %, draw=black, fill=red!30 %fill=white,
	\tikzstyle{format}=[rectangle,draw,thin,align=center,minimum width=1cm,minimum height=0.6cm]%
	\begin{center}
		\begin{tikzpicture}[node distance=2cm,
		auto,>=latex',
		thin,
		start chain=going below,
		every join/.style={norm},]
		\node(start) [startstop] {\small Start};
		\node[below of=start,format,yshift=1.0cm](A){\small Input VOT \& pick a route};
		\node[below of=A,yshift=-0.5cm,circle, draw,thin,fill=white,align=center,text centered,yshift=1.2cm,minimum width=0.2cm,minimum height=0.2cm](B){};
		\node[below of=B,format,yshift=1.2cm](C){\small Reach an intersection};
		\node[draw, diamond, aspect=2,below of=C,text width=2.5cm,align=center,yshift=-0.1cm] (D){\small Are there conflicting vehicles?};
		\node[below of=D,format,fill=green,yshift=-0.15cm](E){\small Determine schedule priority and play a game};
		\node[below of=E,format,fill=green,yshift=0.85cm](F){\small Trade for passing priority};
		\node[below of=F,yshift=0.2cm,circle, draw,thin,fill=white,align=center,text centered,yshift=0.7cm,minimum width=0.2cm,minimum height=0.2cm](G){};
		\node[below of=G,format,yshift=0.85cm](H){\small Reach the destination};
		\node (end) [below of=H,yshift=1cm,startstop] {\small End};
		
		\draw[->] (start.south) -- (A.north); %dotted,
		\draw[dotted,->] (A.south) -- (B.north);
		\draw[->] (B.south) -- (C.north);
		\draw[->] (C.south) -- (D.north);
		\draw[->] (D.south) -- node[anchor=west] {\small yes} (E.north);
		\draw[->] (D.west) -- node[anchor=north]{\small no}(-4,-5.2) -- (-4,-9.6) -- (G.west);
		\draw[->] (E.south) -- (F.north);
		\draw[->] (F.south) -- (G.north);
		\draw[dotted,->] (G.south) -- (H.north);
		\draw[->] (G.east) -- node[anchor=south]{\small go to next intersection}(4,-9.6) -- (4,-2.3) -- (B.east);
		\draw[->] (H.south) -- (end.north);
		%(start.south) edge[dotted,"$w$" '] (A.north);
		\end{tikzpicture}
	\end{center}
	\caption{Flow chart of a trip for all market mechanisms.}\label{F:fc3}
\end{figure}

Competition for priority at an intersection between two vehicles (or vehicle groups) in conflicting approaches can be treated as a game. Rules of the game can assume an auction, direct-transaction, or some other reasonable mechanism. Such a game is illustrated in Fig. \ref{F:fc3}.

%Both game-playing and trading steps need the mechanism's operation. Because intersection passing requires quick-reaction, if humans take place this part of work, congestion and even accidents will happen continually.

\subsection {General analysis}
\label{SS:ga}

For auctions and direct-transactions, travelers only need to provide their VOT compared with travel in a traditional environment. The process is quite simple, easy to learn and implement.

However, credit auctions are much more complicated. Similar to the network credit scheme, the intersection credit scheme should also be hard to attract travelers \cite{krabbenborg2020exploring}. To guarantee that the budget is appropriate for the trip, travelers need to trade frequently, and this requires more time (as part of the trip). If some travelers don't want to incur these additional delays, they have to pay for an increase in budget to guarantee that the credit is enough for travel, and this can result in wasted money because credit will inevitably devalue and expire. Also, trading credit through an agent will result in an additional charge \cite{nie2012transaction}.

%\subsection{Auctions in traffic management}
%\label{SS:auctions}
%
%\subsection{Direct transactions in traffic management}
%\label{SS:transactions}

\section{Analysis of benefits}
\label{S:analysis}

\subsection{Overview and notation}
\label{SS:notation}
The game setting in this paper is one that is played between pairs of vehicles. We consider the simple setting of an intersection of two single-lane one-way streets. That is, we consider two conflicting through movements. In our analysis we take the position of one of the two vehicles involved in a game, which we shall refer to as vehicle A, the \emph{subject vehicle}, and denote their VOT by $v_{\A}$.  We refer to the other vehicle as vehicle B, the \textit{opponent vehicle}, and denote their VOT by $v_{\B}$. Generally speaking, the games considered here are symmetric and it, therefore, does not matter which vehicle we assume to be the subject vehicle and which vehicle we consider to be the opponent. We denote by $t_{\A}$ ($t_{\B}$) the \emph{time saved} by vehicle A (vehicle B) should they win. 

Both the time and money saved are important in the implementation of economic instruments. The VOT is like a bridge used to transfer time into money, and we consider \emph{benefit} $B$ as the sum of money earned and transferred time saved (converted to monetary units). 
For example, if vehicle A with VOT $v_\A = 2$cents/s, gets passing priority and saves $t_\A = 2$s, its time saved in monetary units is $v_\A t_\A = 4$cents. But it pays 1 cent so that its benefit is $B = 4 - 1 = 3$cents. Here, the \emph{return rate of the game} is $\alpha = 3/4 = 0.75$ for vehicle A.
 
 %For vehicle A (vehicle B), we shall refer to $v_{\A} t_{\A}$ ($v_{\B} t_{\B}$) as their \emph{personal benefit}, which is the monetary value they assign to a win.  In an honest game, it reasonable to assume that the players will bid these amounts. 
In brief, we analyze different economic instruments in this section:

1) \emph{First-price auctions}. The higher bid wins and the winner delivers the amount they bid in its entirety to a third party operator in exchange for ROW or priority. Here, the return rate of game $\alpha$ is always zero.

2) \emph{Second-price auctions}. The player with the higher bid also wins, but only delivers the loser's (smaller) bid to the operator. The  $\alpha$  is (this vehicle is A):
\begin{equation}
\alpha_{\mathrm{2auc}} =
\left\{
\begin{array}{ll}
 \frac{v_{\A} t_{\A} - v_\B t_\B}{v_{\A} t_{\A}} & \text{ if A wins}\\
0 &  \text{ if A loses}\\
\end{array}
\right.
\end{equation}

3) \emph{Credit with second-price auctions} (we refer to it as \emph{credit} in this paper). Its benefit from the game is the same as the benefit in the second-price auction ($B_{2auc}$). However, out of game, it still has money earned from credit distribution $C$, and a loss because of credit trading $L_A$. In this paper, we suppose the money earned out of game is  person-equal and revenue-neutral, and its lost is about 10\% of what it earns in total. The credit benefit is:
\begin{equation}
 B_{cre} = B_{2auc} + C - L_A. \label{E:credit}
\end{equation} 

4) \emph{Direct transactions}. The winner transfers a fraction of their winnings to the loser and nothing to a third party operator.  We consider two rules for the value of $\alpha_{\mathrm{tra}}$, the first rule is $\alpha_{\mathrm{tra}} = 1/2$, which has analytical support as a side payment in a transferable utility game \cite{lin2019pay,lin2019transferable}. The second rule is that vehicle A keeps a fraction of their bid equal to the proportion of their personal benefit to the total personal benefit in the system: $\alpha_{\mathrm{tra}} = v_{\A} t_{\A} / (v_{\A} t_{\A} + v_{\B} t_{\B})$.  The two rules can be amalgamated using a binary variable $\chi$, which applies the first rule when set to zero and applies the second rule when set to one:
\begin{equation}
\alpha_{\mathrm{tra}} = \frac{(v_{\A} t_{\A} - 1)\chi + 1}{(v_{\A} t_{\A} + v_{\B} t_{\B} - 2)\chi + 2}.
\end{equation} 

In a sealed bid, vehicles do not know each other's VOTs, so from the perspective of each vehicle, the other vehicle's VOT is a random variable. We will also consider heterogeneity in traffic conditions, so that time savings can also be treated as random variables.  We will quantify and analyze expected net personal benefits and use these expectations to perform comparisons between the different game types.  Use of expected values is motivated by the law of large numbers: it represents the benefits realized after playing a large number of games.  

%Throughout the paper, we will use upper case letters to denote random variables and lower case letters to denote deterministic realizations.  For example, $V_{\A}$ denotes the VOT of A as seen by B, a random quantity, while $v_{\A}$ is A's VOT as seen by A (we assume that vehicle owners know their own VOTs).  We denote by $f_X$ the probability density function of random variable $X$ and $\theta_X$ the parameters of $f_X$; for example $f_{T_{\A}}(t|\theta_{T_{\A}})$ is the probability density at $t$ associated with A's random time savings $T_{\A}$.  We shall drop the dependence on the parameters $\theta_X$ when it is understood from context.  We shall also assume that $V_{\A}$, $V_{\B}$, $T_{\A}$, and $T_{\B}$ are mutually independent random variables, which is a reasonable assumption considering what they represent.  The random variable $T_{\A}$, $T_{\B}$, $V_{\A}$, and $V_{\B}$ all have non-negative support; this fact will be used without mention in our derivations.

%We denote the net personal benefit by $B_{\mathrm{1auc}}$, $B_{\mathrm{2auc}}$, and $B_{\mathrm{tra}}$ for a first-price auction, a second-price auction, and a direct transaction game, respectively.  The functions $B_{\mathrm{1auc}}$, $B_{\mathrm{2auc}}$, and $B_{\mathrm{tra}}$ take the same four arguments: $v_1$ $v_2$, $t_1$, and $t_2$ representing the VOT of the subject vehicle (A or B), the VOT of the opponent, the time saved in a win for the subject vehicle, and  the time saved in a win for the opponent, respectively.

\subsection{Input of calculation}

\begin{figure}[h!]
	\centering
	\resizebox{0.36\textwidth}{!}{%
		\includegraphics{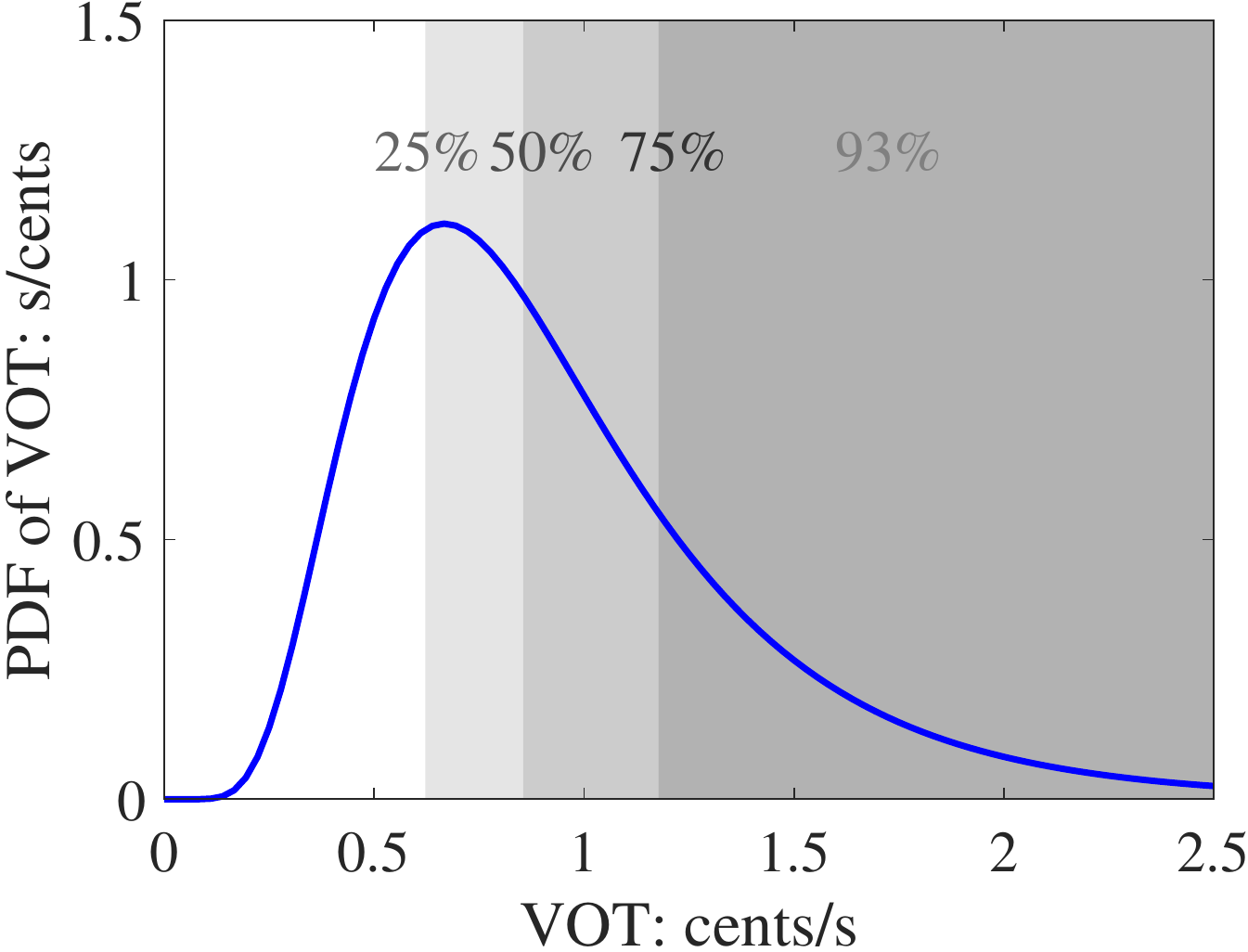}}
	\caption{PDF of VOT.} 
	\label{F:VOT}
\end{figure}

\begin{figure}[h!]
	\centering
	\resizebox{0.36\textwidth}{!}{%
		\includegraphics{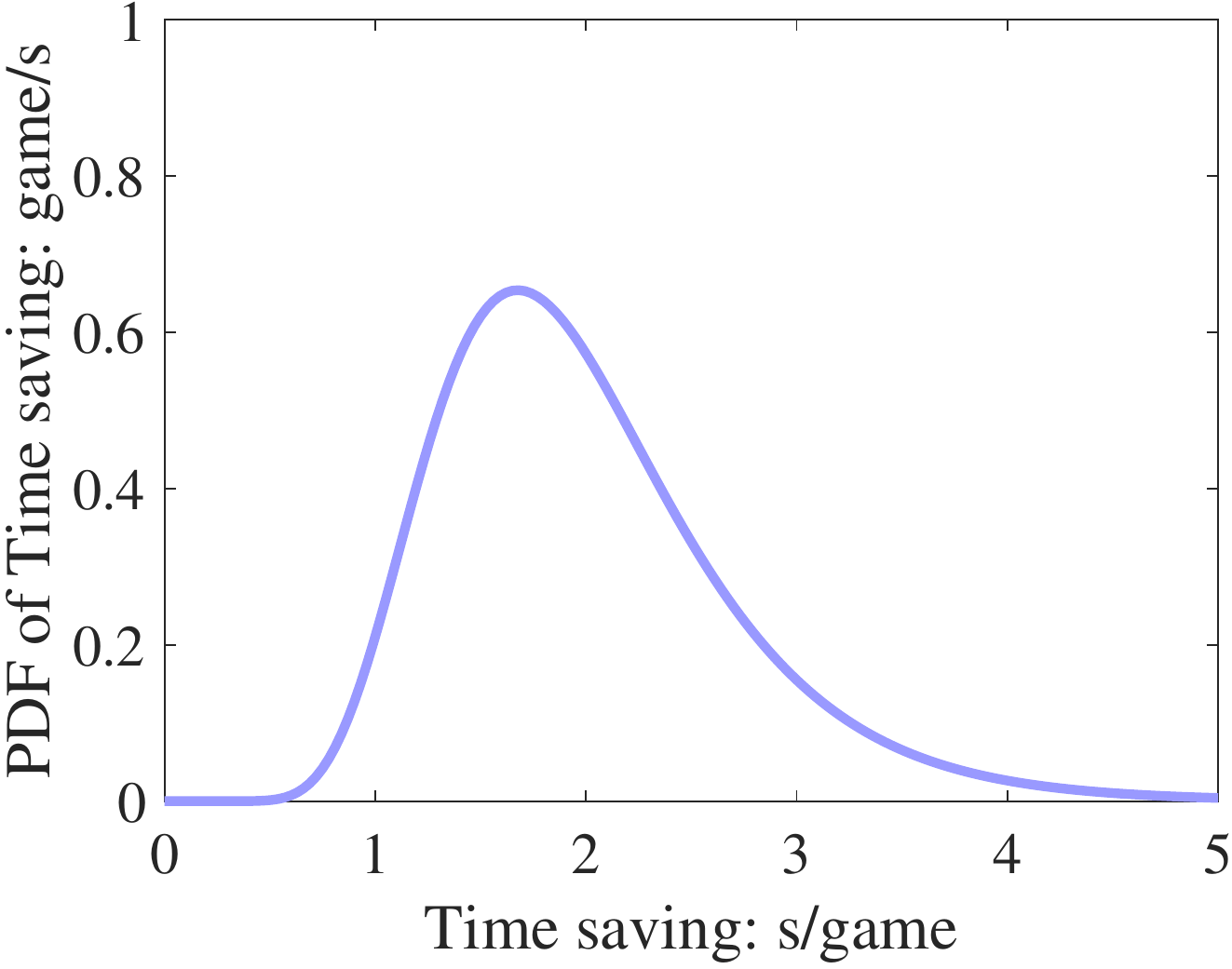}}
	\caption{PDF of time saving.} 
	\label{F:TS}
\end{figure}

%-----------------------
In the calculations below, we assume that the probability distributions of VOT and time saving (across the driving population) are given. In this section, we pick a typical distribution, log-normal distribution for both VOT and time saving. We borrow the investigation results of VOT by Brownstone, et, al in San Diego for $f_v$ \cite{brownstone2003drivers}. As shown in Fig. \ref{F:VOT}, median VOT is $\$30$ per hour (0.8 cent per second). The upper quartile is $\$43$ per hour (1.2 cent per second), and the lower quartile is $\$23$ per hour (0.6 cent per second). In addition, considering a mean discharging time of 2 second per vehicle at the intersection, we design $f_t$ shown as in Fig. \ref{F:TS}.

%-------------------------
\subsection{No abandonment and honest VOT reporting}
\label{SS:nahr}

The net personal benefits of the game are summarized in Table \ref{t_bene_1} for the case when player A wins, and in Table \ref{t_bene_2} for the case when player A loses. Because this is symmetric for A and B, we just focus on vehicle A in the analysis. We define:
\begin{equation}
\beta \equiv 1 -\alpha.
\end{equation}

%In this paper, \emph{net personal benefit} refers to the personal benefit less the amount that is delivered from the winner to the operator (or the loser in the case of a direct transaction game).%\autoref{t_bene_1}
\begin{table}[htbp]%[h!]
	\caption{Benefit of game when vehicle A wins}
	\centering
	\begin{tabular}{|c|c|c|c|}
		\hline
		\multirow{2}{*}{Game} &
		\multicolumn{3}{c|}{Net Benefit} \\
		\cline{2-4}
		& \tabincell{c}{A (winner) }  & \tabincell{c}{B (loser) } &  \tabincell{c}{Operator} \\
		\hline
		First-price auction & 0 & 0 & $v_{\A}t_{\A}$ \\
		Sec.-price auction & $\alpha_{\mathrm{2auc}} v_{\A}t_{\A}$ & 0 & $\beta_{\mathrm{2auc}} v_{\A}t_{\A}$ \\
		Dir. transaction &$\alpha_{\mathrm{tra}} v_{\A}t_{\A}$ &  $\beta_{\mathrm{tra}} v_{\A} t_{\A}$ & 0 \\
		\hline
	\end{tabular}
	\label{t_bene_1}
\end{table}
%In the case of a first-price auction, A delivers the entire amount they bid to the operator and are left with zero net personal benefit.  In a second-price auction, A delivers a fraction $0 < \beta_{\mathrm{auc}} < 1$ of their bid to the operator and are left with $\alpha_{\mathrm{auc}} \equiv 1 - \beta_{\mathrm{auc}}$, which in accord with the discussion above is given by $\beta_{\mathrm{auc}} = v_{\B} t_{\B} / v_{\A} t_{\A}$ (when A wins, i.e., has the higher bid).  Similarly, in a direct transaction game a fraction $0 < \beta_{\mathrm{tra}} < 1$ of A's bid is delivered to B, leaving them with a net personal benefit of $\alpha_{\mathrm{tra}} v_{\A} t_{\A}$, where $\alpha_{\mathrm{tra}} \equiv 1 - \beta_{\mathrm{tra}}$. 
%The net personal benefits to the three parties when B wins the game are listed in \ref{t_bene_2}.  
\begin{table}[h!]
	\caption{Net personal benefits when vehicle B wins}
	\centering
	\begin{tabular}{|c|c|c|c|}
		\hline
		\multirow{2}{*}{Game} &
		\multicolumn{3}{c|}{Net Benefit} \\
		\cline{2-4}
		& \tabincell{c}{A (loser) }  & \tabincell{c}{B (winner) } &  \tabincell{c}{Operator} \\
		\hline
		First-price auction & 0 & 0 & $v_{\B}t_{\B}$ \\
		Sec.-price auction & 0 & $\beta_{\mathrm{2auc}} v_{\B}t_{\B}$ & $\alpha_{\mathrm{2auc}} v_{\B}t_{\B}$ \\
		Dir. transaction &$\alpha_{\mathrm{tra}} v_{\B}t_{\B}$ &  $\beta_{\mathrm{tra}} v_{\B} t_{\B}$ & 0 \\
		\hline
	\end{tabular}
	\label{t_bene_2}
\end{table}

%\subsection{Expected Benefit}
%\label{SS:Ben_one}
%We shall also be interested in the \emph{social benefit}, which is the total benefit to all players involved in the game. In the case of an auction the two players are the operator and either vehicle A or vehicle B; in the case of a direct transaction game, the two players are vehicle A and vehicle B.  We denote the social benefit by $B_{\mathrm{soc}}$; it takes the same arguments that the net personal benefits do and is independent of the type of game, but depends on who wins. If vehicle A wins an auction, they deliver a fraction of their bid to the operator and keep the rest, so that the social benefit is just vehicle A's bid. In a direct transaction game, they deliver a portion of their bid to vehicle B and the social benefit is the total, which is also vehicle A's bid.  Similarly, the social benefit is vehicle B's bid if they win the game, regardless of the type of game.
%
%Our analysis below proceeds along two dimensions; the first is whether users abandon the system or not, the second is whether drivers disclose their true VOTs or not.  Since losers in auction games receive no compensation, it stands to reason that it is possible that drivers with low VOT will not wish to participate in the auctions.  This is our motivation for considering user abandonment. 

%\subsection{No abandonment and honest VOT reporting}
%\label{SS:nahr}
The situation is trivial in this case for a first-price auction: $B_{\mathrm{1auc}} = \EE B_{\mathrm{1auc}} = 0$ with probability 1.  For a second-price auction, vehicle A knows their own VOT, but not B's, so the latter is random.  The time savings shall also be assumed to be random to represent variability in traffic conditions.  From Table \ref{t_bene_1}, the benefit of the game for A in a second-price auction is 
\begin{equation}
B_{\mathrm{2auc}}(v_{\A}, V_{\B}, T_{\A}, T_{\B}) =\bm{1}_{\{ v_{\A} T_{\A} \ge V_{\B} T_{\B} \}} \alpha_{\mathrm{auc}} v_{\A} T_{\A},
\end{equation}
where $\bm{1}_{\{\cdot\}}$ is an indicator function that returns 1 if $\cdot$ is true and returns 0 otherwise.
The expected benefit in the game to A is given by
\begin{multline}
\EE B_{\mathrm{2auc}}(v_{\A}, V_{\B}, T_{\A}, T_{\B}) = \EE  \bm{1}_{\{ v_{\A} T_{\A} \ge V_{\B} T_{\B} \}} \alpha_{\mathrm{auc}} v_{\A} T_{\A} \\ = \EE \big( \alpha_{\mathrm{auc}} v_{\A} T_{\A} \big| v_{\A} T_{\A} \ge V_{\B} T_{\B} \big) \PP(v_{\A} T_{\A} \ge V_{\B} T_{\B}).
\end{multline}

The expected benefit for credit can be based on the expected benefit of second-price auctions in Eq. \ref{E:credit}:
\begin{equation}
\EE B_{\mathrm{cre}} = \EE B_{\mathrm{2auc}} + C - \EE L_A.
\end{equation}

In the case of a direct transaction game, there is a non-negative net personal benefit to vehicle A whether they win or lose.  The net personal benefit is given by
\begin{multline}
B_{\mathrm{tra}}(v_{\A}, V_{\B}, T_{\A}, T_{\B}) = \\ \bm{1}_{\{ v_{\A} T_{\A} \ge V_{\B} T_{\B} \}} \alpha_{\mathrm{tra}} v_{\A} T_{\A} + \bm{1}_{\{ v_{\A} T_{\A} < V_{\B} T_{\B} \}} \alpha_{\mathrm{tra}} v_{\B} T_{\B}
\end{multline}
and the expected personal benefit is
\begin{multline}
\EE B_{\mathrm{tra}}(v_{\A}, V_{\B}, T_{\A}, T_{\B}) = \\\EE \big(\alpha_{\mathrm{tra}} v_{\A} T_{\A} \big| v_{\A} T_{\A} \ge V_{\B} T_{\B} \big) \PP(v_{\A} T_{\A} \ge V_{\B} T_{\B}) + \\ \EE \big( \alpha_{\mathrm{tra}} v_{\B} T_{\B} \big| v_{\A} T_{\A} < V_{\B} T_{\B} \big) \PP(v_{\A} T_{\A} < V_{\B} T_{\B}).
\end{multline}

%To calculate the social benefit, we no longer assume that VOT is known as we do not assume the role of either vehicle. Hence, following the definition above
%\begin{equation}
%B_{\mathrm{soc}}(V_{\A}, V_{\B}, T_{\A}, T_{\B}) = \mathds{1}_{\{ V_{\A} T_{\A} \ge V_{\B} T_{\B} \}} V_{\A} T_{\A} + \mathds{1}_{\{ V_{\A} T_{\A} < V_{\B} T_{\B} \}} V_{\B} T_{\B}
%\end{equation}
%and
%\begin{multline}
%\EE B_{\mathrm{soc}}(V_{\A}, V_{\B}, T_{\A}, T_{\B}) = \EE \big( V_{\A} T_{\A} \big| V_{\A} T_{\A} \ge V_{\B} T_{\B} \big) \PP(V_{\A} T_{\A} \ge V_{\B} T_{\B}) \\+ \EE \big( V_{\B} T_{\B} \big| V_{\A} T_{\A} < V_{\B} T_{\B} \big) \PP(V_{\A} T_{\A} < V_{\B} T_{\B}). \label{E:EBsoc}
%\end{multline}

When a pair of vehicles cross a single-lane one-way intersection, there is a basic scenario that both vehicles have equal opportunity to pass with priority. That is similar to a two-equal-phase control scenario. The Expected benefit in this traditional environment is the \emph{expected basic benefit} $\EE B_b$ in this section. 

We define \emph{expected extra benefit} as any expected benefit minus expected basic benefit. If it is larger than 0, travelers can benefit from the economic instrument. Otherwise, travelers cannot benefit from it. 

\begin{figure}[h!]
	\centering
	\resizebox{0.445\textwidth}{!}{%
		\includegraphics{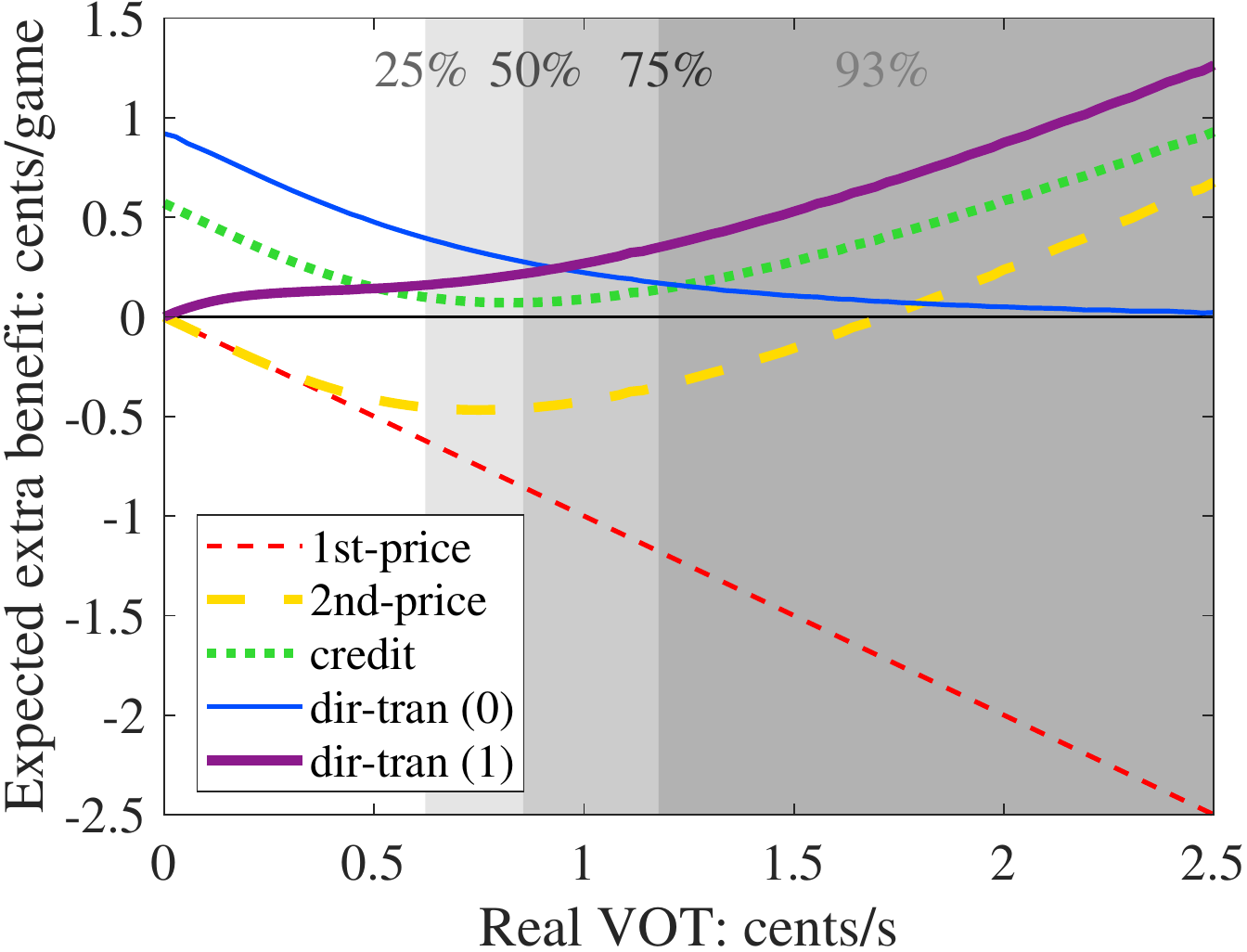}}
	\caption{Benefit in honest scenario.} 
	\label{F:hon}
\end{figure}

Fig. \ref{F:hon} shows the expected extra benefit for different VOT vehicles. The background shows the accumulated distribution of VOT. Obviously, in a first-price auction system, people can gain nothing and the extra benefit in just $-\EE B_b$. They will be indifferent to receiving right-of-way (ROW)  or not. A system that uses the second-price auctions is a system that only benefits the rich. Vehicles with VOTs at and below the median VOT will dislike the auction game and prefer the traditional fairness control. The transaction game with $\chi = 0$  is rewarding the low VOT vehicles. They earn much more in benefit in a transaction game than under traditional control. For these top VOT vehicles, they can still gain some additional benefits. The transaction game with $\chi = 1$ is rewarding to the high VOT vehicles. At the same time, the credit scheme is similar to a balance between the two transaction cases.
 
%-------------------------------
%-------------------------------

\subsection{Honest VOT reporting with abandonment}
\label{SS:ahr}
In this part, we suppose that the traditional traffic system and the market-inspired system will both exist. A vehicle predetermines whether to abandon the market-inspired system before traveling.  To this end, the binary variable $\gamma$ represents whether they choose to play the game ($\gamma = 1$) or not ($\gamma = 0$). If either vehicle does not wish to play the game, then $\gamma = 0$. When this is the case, there is a probability that a vehicle will receive priority and a probability that they will not. Let $p_{\A}$ denote the probability that vehicle A receives priority when $\gamma = 0$ and $p_{\B} \equiv 1 - p_{\A}$. If $\gamma = 0$ and $v_{\A}t_{\A} \ge v_{\B}t_{\B}$, in a first-price auction vehicle A's net personal benefit is $v_{\A}t_{\A}$ with probability $p_{\A}$ and zero with probability $1-p_{\A} = p_{\B}$, so that on average vehicle A's net personal benefit is $p_{\A}v_{\A}t_{\A}$.  Similar arguments can be made for vehicle B, other types of games, and when $v_{\A}t_{\A} < v_{\B}t_{\B}$.  The (mean) net personal benefits of the game are summarized in Tables \ref{t_bene_ul_1} and \ref{t_bene_ul_2}.
\begin{table}[h!]
	\caption{Net personal benefits when $v_{\A}t_{\A} \ge v_{\B}t_{\B}$}
	\centering
	\begin{tabular}{|c|c|c|c|}
		\hline
		\multirow{2}{*}{Game} &
		\multicolumn{3}{c|}{Net Benefit} \\
		\cline{2-4}
		& \tabincell{c}{A (winner)}  & \tabincell{c}{B (loser)} &  \tabincell{c}{Operator} \\
		\hline
		First-price & $(1 - \gamma) p_{\A} v_{\A}t_{\A}$  & $(1 - \gamma)p_{\B}v_{\B}t_{\B}$ & $\gamma v_{\A}t_{\A}$ \\
		\thead{Sec.-price} & \thead{$\gamma \alpha_{\mathrm{2auc}} v_{\A}t_{\A} +$\\$(1-\gamma)p_{\A}v_{\A}t_{\A}$}   & \thead{$(1-\gamma)p_{\B}v_{\B}t_{\B}$} & \thead{$\gamma\beta_{\mathrm{2auc}} v_{\A}t_{\A}$} \\
		\thead{Dir. trans} &\thead{$\gamma \alpha_{\mathrm{tra}} v_{\A}t_{\A}  +$\\$(1-\gamma)p_{\A}v_{\A}t_{\A}$} &  \thead{$\gamma \beta_{\mathrm{tra}} v_{\A} t_{\A} +$\\$(1-\gamma)p_{\B}v_{\B}t_{\B}$} & \thead{0} \\
		\hline
	\end{tabular}
	\label{t_bene_ul_1}
\end{table}

\begin{table}[h!]
	\caption{Net personal benefits when $v_{\A}t_{\A} < v_{\B}t_{\B}$}
	\centering
	\begin{tabular}{|c|c|c|c|}
		\hline
		\multirow{2}{*}{Game} &
		\multicolumn{3}{c|}{Net Benefit} \\
		\cline{2-4}
		& \tabincell{c}{A (loser)}  & \tabincell{c}{B (winner)} &  \tabincell{c}{Operator} \\
		\hline
		First-price & $(1 - \gamma)p_{\A}v_{\A}t_{\A}$ & $(1 - \gamma)p_{\B}v_{\B}t_{\B}$ & $\gamma v_{\B}t_{\B}$ \\
		\thead{Sec.-price} & \thead{$(1 - \gamma)p_{\A}v_{\A}t_{\A}$} & \thead{$\gamma \beta_{\mathrm{2auc}} v_{\B}t_{\B} +$\\ $(1 - \gamma)p_{\B}v_{\B}t_{\B}$} & \thead{$\gamma \alpha_{\mathrm{2auc}} v_{\B}t_{\B}$} \\
		\thead{Dir. trans} &\thead{$\gamma \alpha_{\mathrm{tra}} v_{\B}t_{\B}  +$\\$ (1 - \gamma)p_{\A}v_{\A}t_{\A}$} &  \thead{$\gamma\beta_{\mathrm{tra}} v_{\B} t_{\B} +$\\$ (1 - \gamma)p_{\B}v_{\B}t_{\B}$} & \thead{0} \\
		\hline
	\end{tabular}
	\label{t_bene_ul_2}
\end{table}

%\textbf{First-price auctions}. Notice that when vehicle abandonment is permitted, the net personal benefit associated with a first-price auction can be greater than zero!  When abandonment is allowed, the net personal benefit to vehicle A is
%\begin{equation}
%B_{\mathrm{1auc}}(v_{\A}, V_{\B}, T_{\A}, T_{\B}) = (1-\gamma)p_{\A} v_{\A} T_{\A}.
%\end{equation}
%The expected benefit is readily given by $\EE B_{\mathrm{1auc}}(v_{\A}, V_{\B}, T_{\A}, T_{\B}) = (1-\gamma)p_{\A} v_{\A} \EE T_{\A}$.  When $\gamma = 0$ (vehicle A abandons the system), it is easy to see that $B_{\mathrm{1auc}} \ge 0$ with probability 1, otherwise the net personal benefit is zero with probability 1.  \emph{This suggests that vehicles always have an incentive to abandon first-price auctions.}  
%
%\smallskip
%
%\textbf{Second-price auctions}. In the case of a second-price auction, the net personal benefit to vehicle A is
%\begin{multline}
%B_{\mathrm{2auc}}(v_{\A}, V_{\B}, T_{\A}, T_{\B}) = \\ \mathds{1}_{\{ v_{\A} T_{\A} \ge V_{\B} T_{\B} \}} \big(\gamma \alpha_{\mathrm{auc}}v_{\A} T_{\A} + (1-\gamma)p_{\A} v_{\A} T_{\A} \big) + \\  \mathds{1}_{\{ v_{\A} T_{\A} < V_{\B} T_{\B} \}}(1-\gamma)p_{\A} v_{\A} T_{\A},
%\end{multline}
%which is non-negative whether $\gamma = 0$ or $\gamma = 1$.  

\begin{figure}[h!]
	\centering
	\resizebox{0.445\textwidth}{!}{%
		\includegraphics{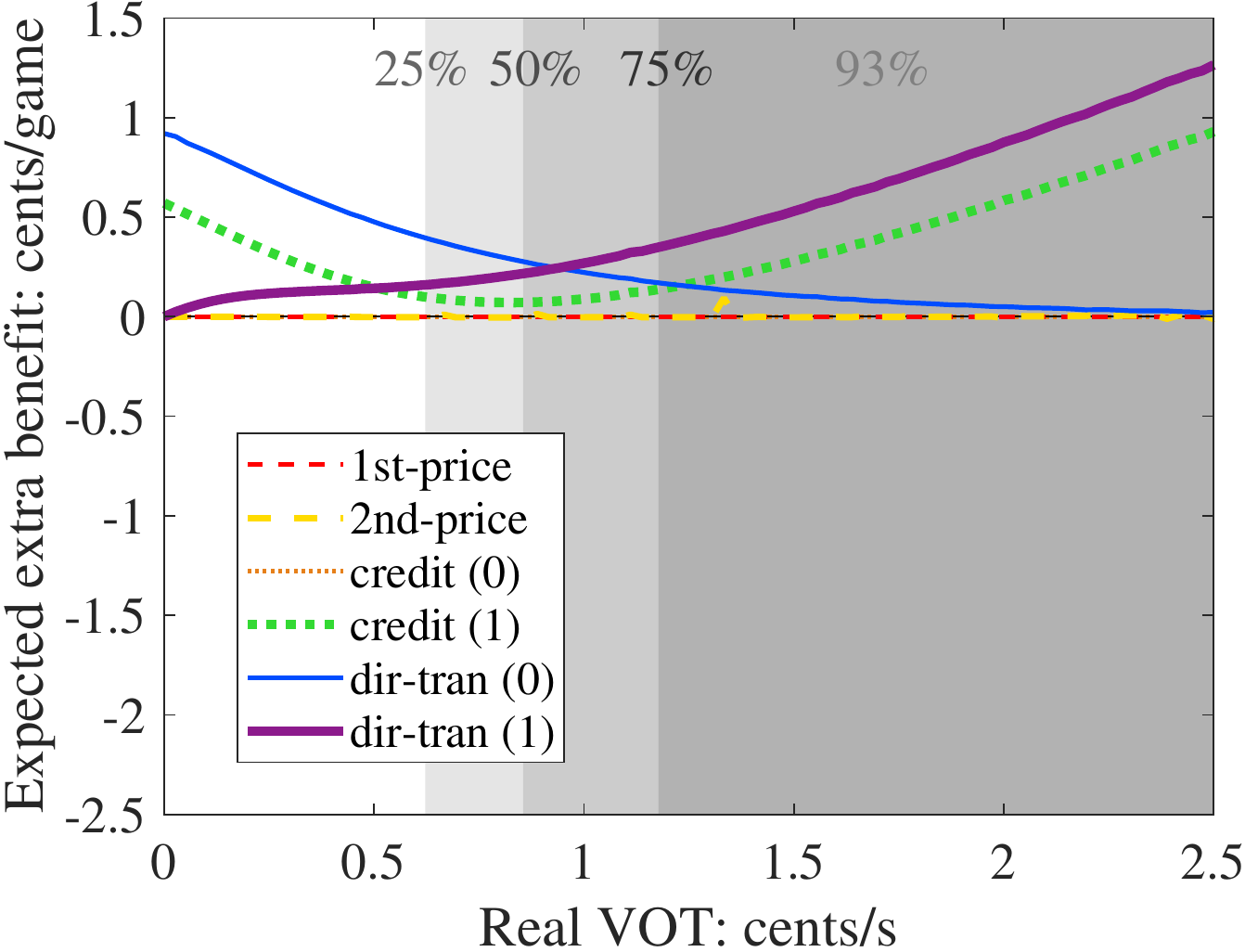}}
	\caption{Benefit in leaving scenario.} 
	\label{F:lea}
\end{figure}

%Limited by pages, we remove expressions of benefits and iteration process in this paper. 
After all travelers' choices are stable, the expected extra benefits will be all equal or larger than zero shown  as in Fig. \ref{F:lea}. We use credit (0) to represent the policy that credits are distributed to all travelers, and credit (1) to represent the policy that credits are only distributed to joiners. Fig. \ref{F:lea} indicates that, under 1st-price auctions, 2nd-price auctions, and normal credit distribution (combined with 2nd-price auctions), all travelers will inevitably abandon the market-mechanism. The problem is that once some travelers find that abandonment can benefit them, they will leave. And then these leavers will lead to further abandonment of the other travelers. Note that credit (1) is much harder to implement and costs a lot for the government because it requires the exact knowledge of traffic users every day.

%-------------------------

\subsection{Dishonest VOT reporting without abandonment}
\label{SS:nadr}
Suppose that the government can only know citizens' declared VOT. All travelers are still rational.
To increase net personal benefits, vehicles/drivers may place bids based on false VOTs.  We denote by $\widetilde{v}_{\A}$ the reported VOT and allow dishonest VOTs to be drawn from different distributions $f_{\widetilde{V}_{\A}}$ for vehicle A and $f_{\widetilde{V}_{\B}}$ for vehicle B. We also use the notation $\widetilde{\alpha}_{\mathrm{auc}}$ and $\widetilde{\alpha}_{\mathrm{tra}}$ to indicate that these fractions are calculated based on reported VOTs.  The net personal benefits realized are still based on true VOT.  For example, if vehicle A bids $\widetilde{v}_{\A} t_{\A}$ in a first-price auction and wins, their net personal benefit is $(v_{\A} - \widetilde{v}_{\A})t_{\A}$. If system abandonment is not allowed, net personal benefits in a system with dishonest VOT reporting are summarized in Table \ref{t_bene_ls_1} when vehicle A wins and in Table \ref{t_bene_ls_2} when vehicle B wins. The declared VOT distribution will keep dynamically changing. If the operator also uses a dynamic economic instrument (for example, $\chi$ is not fixed for the direct transaction), the benefit will also be unstable.

\begin{figure}[h!]
	\centering
	\resizebox{0.445\textwidth}{!}{%
		\includegraphics{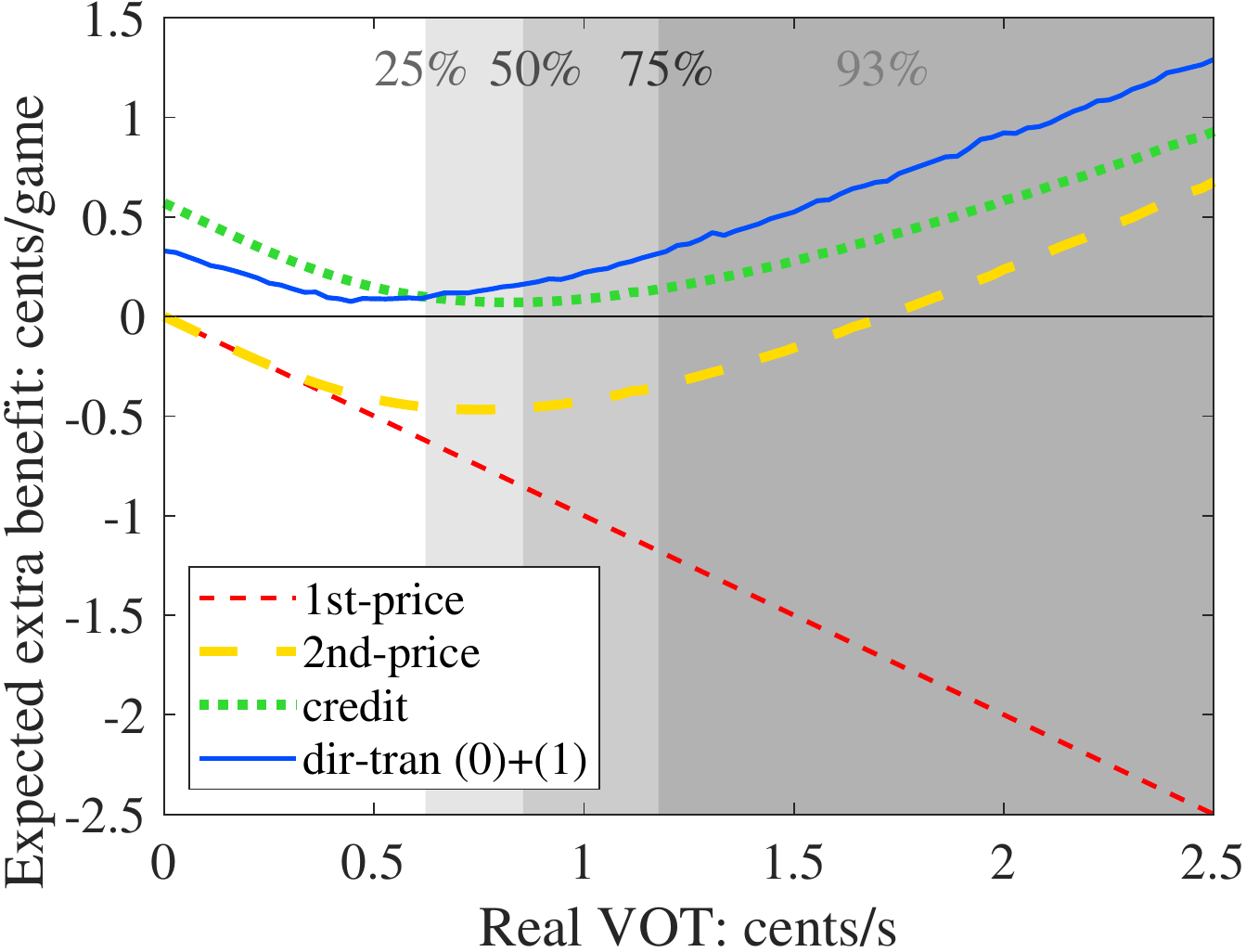}}
	\caption{Benefit in lying scenario.} 
	\label{F:lyi}
\end{figure}

%The (mean) net personal benefits of the game are summarized in Tables \ref{t_bene_ls_1} and \ref{t_bene_ls_2}. 
After about 20 iterations, all expected extra benefits become stable or repeatable. We show their values/mean values in Fig. \ref{F:lyi}. 

\begin{table}[h!]
	\caption{Net personal benefits when $\widetilde{v}_{\A}t_{\A} \ge \widetilde{v}_{\B}t_{\B}$}
	\centering
	\begin{tabular}{|c|c|c|c|}
		\hline
		\multirow{2}{*}{Game} &
		\multicolumn{3}{c|}{Net Benefit} \\
		\cline{2-4}
		& \tabincell{c}{A (winner) }  & \tabincell{c}{B (loser) } &  \tabincell{c}{Operator} \\
		\hline
		First-price & $(v_{\A} - \widetilde{v}_{\A})t_{\A}$ & 0 & $\widetilde{v}_{\A}t_{\A}$ \\
		Sec.-price & $(v_{\A} - \widetilde{\beta}_{\mathrm{2auc}} \widetilde{v}_{\A})t_{\A}$ & 0 & $\widetilde{\beta}_{\mathrm{2auc}} \widetilde{v}_{\A}t_{\A}$ \\
		Dir. trans & $(v_{\A} - \widetilde{\beta}_{\mathrm{tra}} \widetilde{v}_{\A})t_{\A}$ &  $\widetilde{\beta}_{\mathrm{tra}} \widetilde{v}_{\A} t_{\A}$ & 0 \\
		\hline
	\end{tabular}
	\label{t_bene_ls_1}
\end{table}

\begin{table}[h!]
	\caption{Net personal benefits when $\widetilde{v}_{\A}t_{\A} < \widetilde{v}_{\B}t_{\B}$}
	\centering
	\begin{tabular}{|c|c|c|c|}
		\hline
		\multirow{2}{*}{Game} &
		\multicolumn{3}{c|}{Net Benefit} \\
		\cline{2-4}
		& \tabincell{c}{A (loser) }  & \tabincell{c}{B (winner) } &  \tabincell{c}{Operator} \\
		\hline
		First-price & 0 & $(v_{\B} - \widetilde{v}_{\B})t_{\B}$ & $\widetilde{v}_{\B}t_{\B}$ \\
		Sec.-price & 0 & $(v_{\B} - \widetilde{\alpha}_{\mathrm{2auc}} \widetilde{v}_{\B})t_{\B}$ & $\widetilde{\alpha}_{\mathrm{2auc}} \widetilde{v}_{\B}t_{\B}$ \\
		Dir. trans& $\widetilde{\alpha}_{\mathrm{tra}} \widetilde{v}_{\B}t_{\B}$ &  $(v_{\B} - \widetilde{\alpha}_{\mathrm{tra}} \widetilde{v}_{\B})t_{\B}$ & 0 \\
		\hline
	\end{tabular}
	\label{t_bene_ls_2}
\end{table}

%\textbf{First-price auction.} The expected net personal benefit to vehicle A is given by
%\begin{multline}
%\EE B_{\mathrm{1auc}}(\widetilde{v}_{\A}, \widetilde{V}_{\B}, T_{\A}, T_{\B}) = \EE \mathds{1}_{\{ \widetilde{v}_{\A} T_{\A} \ge \widetilde{V}_{\B} T_{\B} \}} (v_{\A} - \widetilde{v}_{\A})T_{\A} 
%\\ =  (v_{\A} - \widetilde{v}_{\A}) \EE \big( T_{\A} \big| \widetilde{v}_{\A} T_{\A} \ge \widetilde{V}_{\B} T_{\B}\big) \PP(\widetilde{v}_{\A} T_{\A} \ge \widetilde{V}_{\B} T_{\B}),
%\end{multline}
%which can be made strictly positive with an appropriate choice of $ \widetilde{v}_{\A}$. In the case of honest reporting, the expected net personal benefit is zero with probability one. \emph{It follows that there is clear incentive to be dishonest in the case of first-price auctions.}

It is clear that both credit and transaction games perform well, but pure auction strategies can not make all travelers happy.
%-------------------------
%-------------------------------

\section{Conclusion}
\label{S:Conc}
With the popularity of connected vehicle technology and mobile payment, economic instruments can better serve heterogeneous vehicles with different values of time (VOT) in intersection operations. Most papers consider the influence of economic instruments from the perspective of system optimization, but few papers analyze what the users' concerns are. When a new management method appears, users want it to be easy to learn and operate, that it can save time or money, and that it can reduce the rich-poor gap. We haven't analyzed the rich-poor gap in this paper but propose to do so in future research. We found that auction mechanisms will enlarge the rich-poor difference, and the other mechanisms cannot guarantee a reduction of the gap.

Based on a comparative analysis, we found that the credit scheme and free-market scheme can both benefit users. The second price auction can only benefit high VOT vehicles. However, considering the operational difficulties, a credit scheme is not easy to learn and operate for travelers.

\bibliographystyle{IEEEtran}
\bibliography{refs}

\end{document}